\documentclass[10pt]{article}
\usepackage{graphicx}
\title{Newton's Third Law in the Framework of Special Relativity for Charged Bodies Part 3: Time Dependent Engines}

\author{Asher Yahalom$^{1,2}$ and Moshe Sagi$^3$\\
$^{1}$  Ariel University, Faculty of Engineering,\\
 Department of Electrical \& Electronic Engineering,\\
 Ariel 40700, Israel; asya@ariel.ac.il\\
$^{2}$  Ariel University, Center for Astrophysics, Geophysics,\\
 and Space Sciences (AGASS), Ariel 40700, Israel;\\
 $^{3}$  Holon Institute of Technology, Faculty of  Electrical\\
  \& Electronic Engineering, Holon 5810201, Israel;}

\begin{document}
\maketitle

\begin{abstract}
The Lorentz symmetry group entails physical equations whose solutions are retarded. This leads to the concept of a relativistic engine resulting from retardation of electromagnetic fields. We have shown that Newt\-on'n third law cannot strictly hold in a distributed system, where the different parts are at a finite distance from each other and thus force imbalance is created at the system's center of mass. As the system is affected by a total force for a finite period, mechanical momentum and energy are acquired by the system. In previous works we relied on the fact that the bodies were macroscopically natural. Lately we relaxed this assumption and studied charged bodies, thus analyzing the consequences on a possible electric relativistic engine. On the first paper on this subject we investigated this phenomena in general but gave an example of a system only at the stage of reaching a stationary state, a second paper was devoted to the same but on the atomic scale, here we shall analyze the charge relativistic engine in a general time dependent setting.
\end{abstract}

\newcommand{\beq} {\begin{equation}}
\newcommand{\enq} {\end{equation}}
\newcommand{\ber} {\begin {eqnarray}}
\newcommand{\enr} {\end {eqnarray}}
\newcommand{\eq} {equation}
\newcommand{\eqs} {equations }
\newcommand{\mn}  {{\mu \nu}}
\newcommand{\abp}  {{\alpha \beta}}
\newcommand{\ab}  {{\alpha \beta}}
\newcommand{\sn}  {{\sigma \nu}}
\newcommand{\rhm}  {{\rho \mu}}
\newcommand{\sr}  {{\sigma \rho}}
\newcommand{\bh}  {{\bar h}}
\newcommand{\br}  {{\bar r}}
\newcommand {\er}[1] {equation (\ref{#1}) }
\newcommand {\ern}[1] {equation (\ref{#1})}
\newcommand {\Ern}[1] {Equation (\ref{#1})}
\newcommand{\hdz}  {\frac{1}{2} \Delta z}
\newcommand{\curl}[1]{\vec{\nabla} \times #1} % for curl

\section{Introduction}

Special relativity which is closely connected to the Lorentz symmetry group, was introduced in Einstein's famous 1905 paper: "On the Electrodynamics of Moving Bodies" \cite{Einstein}. This theory was a consequence of observations and the laws of electromagnetism, which were formulated in the nineteenth century by Maxwell in his famous equations \cite{Maxwell,Jackson,Feynman} which owe their present form to Oliver Heaviside \cite{Heaviside}.
A consequence of Maxwell's equations is that an electromagnetic signal travels at the speed of light $c$, and soon it was realized that light is an electromagnetic wave. Later Einstein \cite{Einstein,Jackson,Feynman} formulated the special theory of relativity, which postulates that the speed of light in vacuum $c$ is the maximal allowed speed. According to a prevalent interpretation of relativity, any object, signal or field can not travel faster than the speed of light in vacuum. Hence the phenomena of retardation, if at a distance $R$ from an observer a change is made, the observer will be ignorant of it for a time greater than $\frac{R}{c}$. Thus action and reaction cannot be generated simultaneously because of the signal speed.

Here we mention that the phenomena of retardation exist only in Lorentzian space-times \cite{Primordial,JMP}, those space-times are the only solutions for the case of empty or low density environments \cite{Yahaloma,Yahalomb}. However, in a limited region of space-time such as close to the big-bang \cite{Primordial} an Euclidean metric is possible, which does not enable retardation as is evident in the extreme homogeneity of the cosmic microwave background.

Newton's laws of motion laid the foundation for classical mechanics. The three laws of motion were first compiled by Isaac Newton in his Philosophiae Naturalis Principia Mathematica (Mathematical Principles of Natural Philosophy), first published in 1687 \cite{Newton,Goldstein}. Although the
first law is already mentioned in Philosophiae Principia by Descartes which was published in 1644.
Here we shall be interested initially in the third law, which states: When one body exerts a force on a second body, the second body \textbf{simultaneously} exerts a force equal in magnitude and opposite in direction on the first body.

It follows from Newton's third law, that the total sum of forces in a system which is not affected by external forces is null. This law has a significant number of experimental verifications and is thus one of the corner stones of physical sciences. However, it is easy to see that action and its reaction cannot be generated at exactly the same time because the speed of signal propagation is not infinite. Hence the third law cannot be correct in an exact sense, although it can be assumed valid for most practical applications due to high velocity of signal propagation. Thus the total sum of  forces cannot be zero at every given time. Thus according to Newton's second law: "Force is equal to the rate of change of momentum. For a constant mass, force equals mass times acceleration". A body must gain momentum, in the sense that it's center of mass must be moving.  Moreover, Newton's first law states: "every object will remain at rest or in uniform motion in a straight line unless compelled to change its state by the action of an external force". However, as the body gains momenta due to the retardation phenomena it cannot continue to move on a straight line, hence the first law cannot be correct in an exact sense but only in an approximate sense.

In prevalent systems the above described motion is minute, but perhaps there is a way to make this motion significant, this line of thought lead to the concept of a relativistic engine.

Current locomotive systems are based on material parts; each gains momentum that is equal and opposite to the momentum of the other. A generic example is a rocket that sheds gas to move itself forward. However, retardation effects suggest a novel type of engine which is not composed of two material parts but of field and matter. Forgetting about the field, the material body seems to obtain momentum thus the total momentum increases violating the law of momentum conservation. However, it can be shown that the opposite amount of momentum is attributed to the field \cite{MTAY4}, thus the total momentum is indeed conserved. This is a result of Noether's theorem which dictates that a system that possesses translational symmetry will conserve momentum. The total physical system composed of matter and field is indeed invariant under translations, while every part of the system (either matter or field) is not. Feynman \cite{Feynman} depicted two moving charges, apparently contradicting Newton's third law as forces that the charges induce do not cancel (last part of 26-2), this is resolved in (27-6) in which it is noticed that the momentum gained by the two charge system is removed from the field momentum.

A relativistic engine is thus defined as a system in which its material center of mass is in motion due to the interaction of its {\bf material}
parts. Those  may move with respect to each other or held in a rigid frame. This has no effect as we are interested in the motion of the center of mass only. We emphasize that a relativistic motor allows 3-axis motion (vertical motion included), it does not contain moving parts, it does not consume fuel (and does not emit carbon) it only consumes electromagnetic energy  which may be supplied by solar panels or batteries. The relativistic engine is a perfect solution for space travel in which much of the space vehicle volume is devoted to fuel storage.

 Griffiths \&  Heald \cite{Griffiths}  pointed out that the laws of Coulomb's and Biot-Savart determine the electric and magnetic field configurations solely for static sources. Time-dependent generalizations of these laws described by Jefimenko \cite{Jefimenko} were used to study the applicability of Coulomb and Biot-Savart formulas outside the static domain.
In an earlier paper, we made use of Jefimenko's \cite{Jefimenko,Jackson} equation to study the force developing between two current loops \cite{MTAY1}. This was later generalized to include the forces between a current carrying loop and a permanent magnet  \cite{MTAY3,AY1}. Since the device is forced for a finite period, it will gain mechanical momentum and energy.  The subject of momentum conversation was discussed in \cite{MTAY4}. In \cite{AY2,RY,RY2,RY3} the exchange of energy between the mechanical part of the relativistic engine and the electromagnetic field were discussed. In particular, it was shown that the total electromagnetic energy expenditure is six times the kinetic energy gained by the relativistic motor. It was also shown how some energy might be radiated from the relativistic engine device if the coils are not configured properly.

Previous works relied on the fact that the bodies were macroscopically natural, which means that the number of electrons and ions is equal in every volume element. In a later work \cite{RY4} we relaxed this assumption and studied charged bodies, thus analysing the consequences of charge on a possible electric relativistic engine. Notice, however, that in the previous paper we only considered a relativistic engine reaching a stationary state while ignoring the possibility of engines which
do not reach a stationary state as well as the transient stage of engines that do. The same criticism holds for a the second part of the same paper dealing with a relativistic engine on the atomic scale \cite{nano}. Here we make a more general analysis leading to a different kind of charged relativistic engine of a type that does not reach a stationary state.

\section{Generated Momentum}

According to Newton's second law, a system with a non zero total force in its center of mass, must have a change in its total linear momentum $\vec P (t)$:
\beq
\vec F_{T}^{[2]} = \frac{d \vec P}{dt}
\label{P}
\enq
Assuming that $\vec P (-\infty) = 0$ and also that there are not current or charge densities at $t=-\infty$, it follows from equation (81) of \cite{RY4} that:
\beq
\vec P (t)= \frac{\mu_0}{4 \pi}\int \int d^3 x_1  d^3 x_2 ~
  \left[\frac{1}{2}\left(\rho_2 \partial_t \rho_1 - \rho_1 \partial_t \rho_2 \right)\hat R - (\rho_1  \vec J_2 + \rho_2  \vec J_1)  R^{-1}  \right]
\label{P2}
\enq
Comparing \ern{P2} with the momentum gain of a non charged relativistic motor described by equation (64) of \cite{MTAY4}:
\beq
\vec P_{mech} \cong \frac{\mu_0}{8 \pi} \partial_t I_1 (t) I_2 (\frac{h}{c})^2  \vec K_{122}, \qquad  \vec K_{122}=
- \frac{1}{h^2}  \oint \oint \hat R (d\vec l_2 \cdot d\vec l_1)
\label{Pmech1b}
\enq
where $h$ is a typical scale of the system, $I_1 (t)$ and  $I_2$ are the currents flowing through two current loops and
$d\vec l_1, d\vec l_2$ are the current loop line elements. We notice some major differences. First,  we notice a factor of $\left(\frac{h}{c}\right)^2$ in case of an uncharged motor. As, for any practical system the scale $h$ is of the order of one, this means that the charged relativistic motor is stronger than the uncharged motor by a factor of $c^2 \sim 10^{17}$ which is quite a considerable factor. Second, we notice that for the uncharged motor, the current must be continuously increased in order to maintain the momentum in the same direction. Of course, one cannot do this for ever, hence the uncharged motor is a type of a piston engine doing a periodic motion backward and forward and can only produce motion forward by interacting with an external system (the road). This is not the case for the charged relativistic motor. In fact we obtain non vanishing momentum for stationary charge and current densities:
\beq
\vec P (t)= - \frac{\mu_0}{4 \pi} \int \int d^3 x_1  d^3 x_2 ~ (\rho_1  \vec J_2 + \rho_2  \vec J_1)  R^{-1}
\label{P3}
\enq
hence the charged relativistic motor can produce forward momentum without interacting with any external system except the electromagnetic field.

\section {Engine Optimization}

At this stage we would like to investigate what are the conditions for a system in terms of composition and structure to generate a maximal amount of momentum. To this end we write the charge
density and current density in the form:
\beq
\rho (\vec x,t) = \bar \rho \sum_n \rho^n (\vec x)  f^n (t)
\label{rhoexp}
\enq
in which $\bar \rho$ is a constant which has the units of current density, $\rho^n (\vec x)$ are spatial functions and $f^n (t)$ temporal functions, both functions are dimensionless. Similarly:
\beq
\vec J (\vec x,t) = \bar J \sum_n \vec J^n (\vec x)  g^n (t)
\label{Jexp}
\enq
in which $\bar J$ is a constant which has the units of current density, $\vec J^n (\vec x)$ are spatial vector functions and $g^n (t)$ temporal functions, both functions are dimensionless. We shall write the dimensional constants in terms of a generic charge $Q$, typical length scale $h$, and typical time scale $\tau$, that is:
\beq
\bar \rho \equiv \frac{Q}{h^3}, \qquad \bar J \equiv \bar \rho \frac{h}{\tau} = \frac{Q}{h^2 \tau}.
\label{consdim}
\enq
Next we shall attribute an expansion of the type given in \ern{rhoexp} and \ern{Jexp} to each of the
subsystems introduced previously this is done by adding a subscript with the relevant subsystem number:
\beq
\rho_1 (\vec x,t) = \bar \rho_1 \sum_n \rho_1^n (\vec x)  f^n (t), \qquad
\rho_2 (\vec x,t) = \bar \rho_2 \sum_n \rho_2^n (\vec x)  f^n (t).
\label{rhoexpsi}
\enq
\beq
\vec J_1 (\vec x,t) = \bar J_1 \sum_n \vec J_1^n (\vec x)  g^n (t), \qquad
\vec J_2 (\vec x,t) = \bar J_2 \sum_n \vec J_2^n (\vec x)  g^n (t).
\label{Jexpsi}
\enq
Next we define the following dimensionless vector constants:
\beq
\vec \Gamma^{mn} \equiv \vec \Gamma_{21}^{mn} \equiv
\frac{1}{2 h^6} \int d^3 x_1 d^3 x_2 \rho_1^n \rho_2^m \frac{\vec x_1 - \vec x_2}{|\vec x_1 - \vec x_2|} = - \vec \Gamma_{12}^{nm}
\label{Gam}
\enq
which depend on the spatial structure of the two charged systems. Similarly we define:
\beq
 \vec \Lambda_{21}^{mn} \equiv
\frac{1}{h^5} \int d^3 x_1 d^3 x_2 \frac{\rho_2^m \vec J_1^n}{|\vec x_1 - \vec x_2|},
\qquad
\vec \Lambda_{12}^{mn} \equiv
\frac{1}{h^5} \int d^3 x_1 d^3 x_2 \frac{\rho_1^m \vec J_2^n}{|\vec x_1 - \vec x_2|},
\label{Lam}
\enq
we notice that generically:
\beq
\vec \Lambda_{21}^{mn} \neq \vec \Lambda_{12}^{mn}, \qquad \vec \Lambda_{21}^{mn} \neq \vec \Lambda_{12}^{mn}.
\label{Lamineq}
\enq
Using the typical time scales of each of the systems $\tau_1$ and $\tau_2$ we define:
\beq
 \vec \Lambda^{mn} \equiv \frac{1}{\tau_1 +\tau_2 }
 \left(\tau_2 \vec \Lambda_{21}^{mn} + \tau_1 \vec \Lambda_{12}^{mn}\right),
 \qquad \bar \tau = \frac{\tau_1 \tau_2}{\tau_1 +\tau_2}.
\label{Lamtau}
\enq
Plugging \ern{rhoexpsi} and \ern{Jexpsi} into \ern{P2} and using the definitions above we arrive at the expression:
\beq
\vec P (t) = \frac{\mu_0}{4 \pi} Q_1 Q_2 \sum_{mn} \left[\vec \Gamma^{mn} (f^m \partial_t f^n
- f^n \partial_t f^m) - \frac{1}{\bar \tau} \vec \Lambda^{mn} f^m g^n \right].
\label{momgf}
\enq
We remind the reader that $\rho$ and $\vec J$ are not independent as they are connected through the continuity equation (16) of \cite{RY4}:
\beq
\vec \nabla \cdot \vec J + \partial_t \rho = 0 \Rightarrow \sum_n \left[ \bar \rho \rho^n (\vec x)  \partial_t f^n (t) + \bar J \vec \nabla \cdot \vec J^n g^n (t) \right] = 0.
\label{Chacon2}
\enq
Taking into account \ern{consdim} it follows that:
\beq
\sum_n \left[ \rho^n (\vec x) \tau  \partial_t f^n (t) + h \vec \nabla \cdot \vec J^n (\vec x) g^n (t) \right] = 0.
\label{Chacon3}
\enq
As we can always choose:
\beq
g^n (t) = \tau  \partial_t f^n (t),
\label{gchoi}
\enq
the following equation must be satisfied:
\beq
\sum_n \left[ (\rho^n (\vec x)+ h \vec \nabla \cdot \vec J^n (\vec x)) g^n (t) \right] = 0.
\label{Chacon4}
\enq
One possible solution is by choosing:
\beq
\rho^n (\vec x)= - h \vec \nabla \cdot \vec J^n (\vec x),
\label{Chacon5}
\enq
for all $n$. In this case we can choose $\vec J^n$ but $\rho^n$ are dictated.

\section{General Considerations}

Looking at the momentum \ern{momgf} it is obvious that the higher the charge the higher the momentum, also a short time scale $\bar \tau$ and a high time derivative will also increase momentum. Counter intuitively a proximity of charge and currents is also a contributing factor, through the lambda terms given in \ern{Lam} this can be explained due to the fact that interaction
is stronger in close proximity despite the fact that retardation is smaller. In fact interaction increases in proportion to the inverse of the distance square as the systems become closer, however, retardation only decreases linearly in the distance. We remind the reader that in a previous
paper \cite{RY4} we have shown that the amount of charge in a given volume is limited due to the phenomenon of dielectric breakdown, hence to achieve high momentum one needs high charge and thus a large engine as expected.

\section{Example}

\begin{figure}
\centering
\includegraphics[width=0.7\columnwidth]{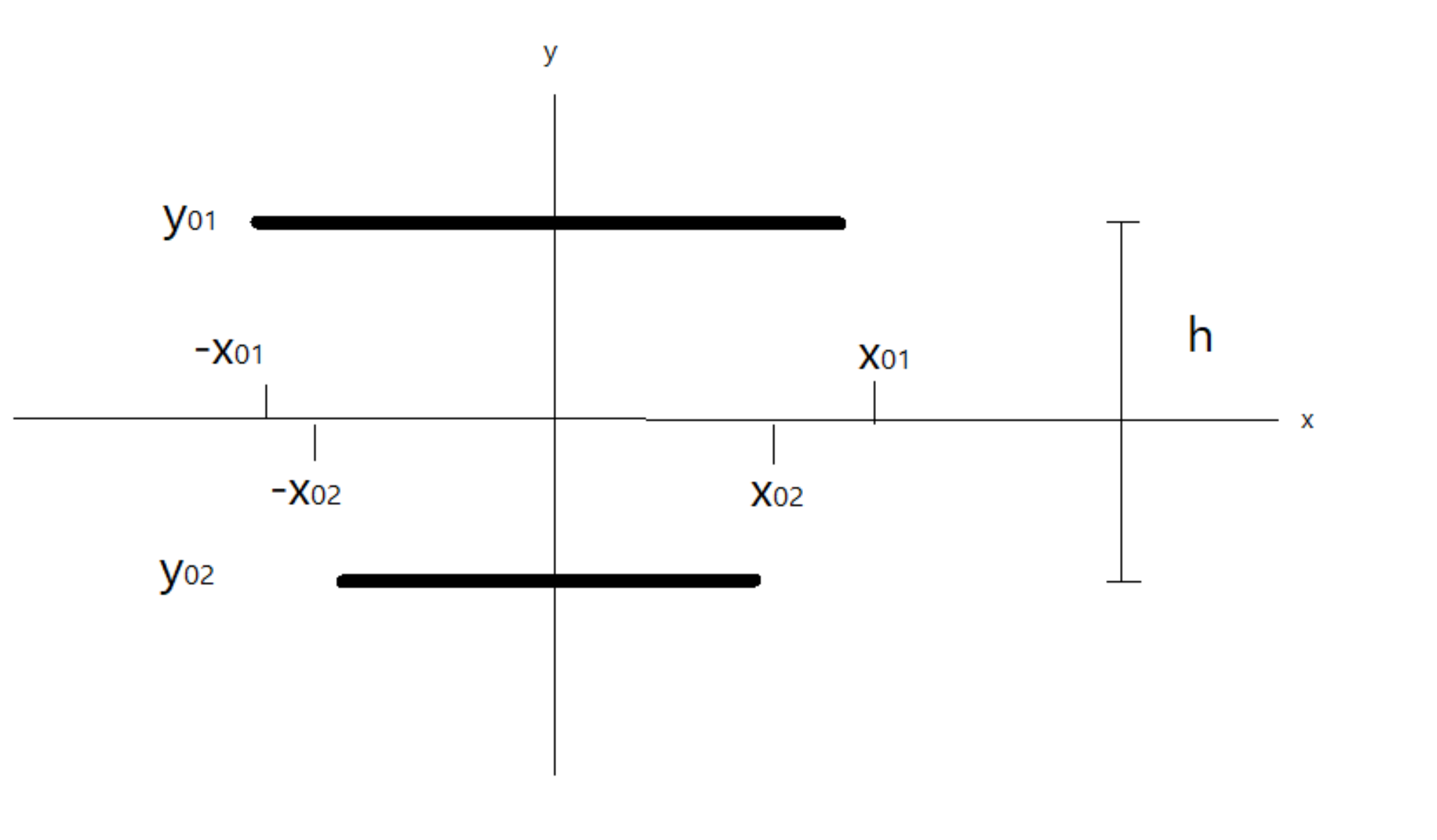}
\caption {Two current strips.}
 \label{Twostrip}
\end{figure}

Let us consider a set of two functions:
\beq
f^1 (t) = \sin \phi_1 (t), \qquad f^2 (t) = \sin \phi_2 (t)
\label{f12}
\enq
in which $\phi_1 (t)$ and $\phi_2 (t)$ are arbitrary time dependent phases. We use \ern{gchoi}
to choose corresponding $g$ functions:
\beq
g^1 (t) = \tau \dot \phi_1 (t) \cos \phi_1 (t), \qquad
g^2 (t) = \tau \dot \phi_2 (t) \cos \phi_2 (t).
\label{g12}
\enq
We shall assume two sub systems each described by only one component in the sum:
\beq
\vec J_1 (\vec x,t) = \bar J_1  \vec J_1^1 (\vec x)  g^1 (t), \qquad
\vec J_2 (\vec x,t) = \bar J_2 \vec J_2^2 (\vec x)  g^2 (t).
\label{Jexpsiex}
\enq
Taking into account \ern{Chacon5} it follows that:
\beq
\rho_1 (\vec x,t) = \bar \rho_1  \rho_1^1 (\vec x)  f^1 (t), \qquad
\rho_2 (\vec x,t) = \bar \rho_2  \rho_2^2 (\vec x)  f^2 (t).
\label{rhoexpsiex}
\enq
Only the following terms in \ern{Gam} and \ern{Lam} are non zero:
\beq
\vec \Gamma^{21} = \frac{1}{2 h^6} \int d^3 x_1 d^3 x_2 \rho_1^1 \rho_2^2 \frac{\vec x_1 - \vec x_2}{|\vec x_1 - \vec x_2|},
\label{Gam21}
\enq
\beq
 \vec \Lambda_{21}^{21} \equiv
\frac{1}{h^5} \int d^3 x_1 d^3 x_2 \frac{\rho_2^2 \vec J_1^1}{|\vec x_1 - \vec x_2|},
\qquad
\vec \Lambda_{12}^{12} \equiv
\frac{1}{h^5} \int d^3 x_1 d^3 x_2 \frac{\rho_1^1 \vec J_2^2}{|\vec x_1 - \vec x_2|}.
\label{Lam12}
\enq
Assuming $\tau_1 = \tau_2 $ it follows from \ern{Lamtau} that $\bar \tau = \frac{\tau_1}{2}$,
we shall take $\tau = \bar \tau$. Also it follows from \ern{Lamtau} that:
\beq
 \vec \Lambda^{12}  = \frac{1}{2} \vec \Lambda_{12}^{12}, \qquad
 \vec \Lambda^{21}  = \frac{1}{2} \vec \Lambda_{21}^{21}.
\label{Lam12b}
\enq
Inserting the above results in \ern{momgf} will yield:
\beq
\vec P (t) = \frac{\mu_0}{4 \pi} Q_1 Q_2  \left[\vec \Gamma^{21} (f^2 \partial_t f^1
- f^1 \partial_t f^2) - \frac{1}{\bar \tau} \vec \Lambda^{21} f^2 g^1
- \frac{1}{\bar \tau} \vec \Lambda^{12} f^1 g^2 \right].
\label{momgfex}
\enq
Let us make the following assumptions. First: we assume that the phases differ by a constant value
$\Delta \phi$. This means that the derivatives of the phases are the same. Taking into account \ern{f12} and \ern{g12} we thus obtain:
\beq
\vec P (t) = \frac{\mu_0}{4 \pi} Q_1 Q_2 \dot \phi  \left[\vec \Gamma^{21} \sin \Delta \phi - \vec \Lambda^{21} \sin \phi_2 \cos \phi_1 - \vec \Lambda^{12} \sin \phi_1 \cos \phi_2 \right].
\label{momgfex2}
\enq
The $\Gamma$ and $\Lambda$ constants are determined by the distribution of charges and currents. Let us consider two thin wires of current as depicted in figure \ref{Twostrip}.
In this case we obtain:
\ber
\vec J_1^1 &=& h^2 \delta (z_1) \delta (y_1-y_{01})\left[u(x_1+x_{01})-u(x_1-x_{01})\right] \hat x,
\nonumber \\
\vec J_2^2 &=& h^2 \delta (z_2) \delta (y_2-y_{02})\left[u(x_2+x_{02})-u(x_2-x_{02})\right] \hat x.
\label{J11J22}
\enr
in which $\delta$ is Dirac's delta function and $u$ is a step function, $\hat x$ is a unit vector in the $x$ direction. It follows from \ern{Chacon5} that we also have:
\ber
\rho_1^1 &=& h^3 \delta (z_1) \delta (y_1-y_{01})\left[\delta(x_1-x_{01})-\delta(x_1+x_{01})\right],
\nonumber \\
\rho_2^2 &=& h^3 \delta (z_2) \delta (y_2-y_{02})\left[\delta(x_2-x_{02})-\delta(x_2+x_{02})\right].
\label{rh11rh22}
\enr
we shall take the characteristic length of the system to be the distance between the two wires
(see figure \ref{Twostrip}):
\beq
h = y_{02} - y_{01}.
\label{hdef}
\enq
Now, we can plug \ern{rh11rh22} into \ern{Gam21} and obtain:
\beq
\vec \Gamma^{21} = h \hat y \left\{ \frac{1}{\sqrt{(x_{01}+x_{02})^2 + h^2}}
- \frac{1}{\sqrt{(x_{01}-x_{02})^2 + h^2}} \right\},
\label{Gam21b}
\enq
For the simple case: of $x_{01} = x_{02}$, it follows that for a wire of length $L = 2 x_{01}
= x_{01} + x_{02}$, it takes the form:
\beq
\vec \Gamma^{21} =  \hat y \left\{ \frac{1}{(\sqrt{(\frac{L}{h})^2 + 1}}.
- 1 \right\},
\label{Gam21c}
\enq
Thus if the wires is very long with respect to the distance between wires, $L \gg h$ it follows that:
\beq
\vec \Gamma^{21} = - \hat y.
\label{Gam21d}
\enq
For a vanishing short wire $\vec \Gamma^{21} = 0$, while for a short wire $L \ll h$, we have:
 \beq
\vec \Gamma^{21} = - \frac{1}{2} (\frac{L}{h})^2 \hat y.
\label{Gam21e}
\enq
The absolute value of $|\Gamma^{21}|$ is depicted in figure \ref{gamf}:
\begin{figure}
\centering
\includegraphics[width=0.7\columnwidth]{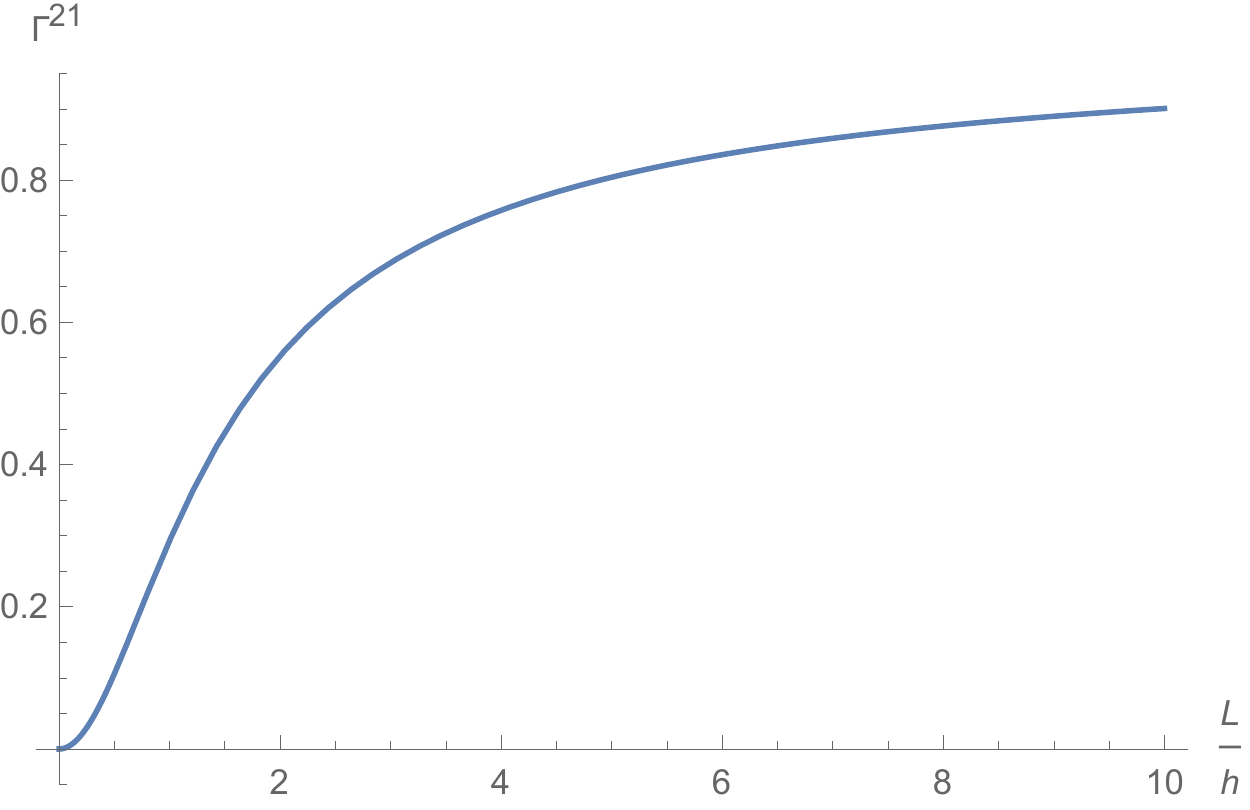}
\caption {$|\Gamma^{21}|$, the functions approached asymptotically to a unit value.}
 \label{gamf}
\end{figure}
Let us now insert \ern{J11J22} and \ern{rh11rh22} into \ern{Lam12} and \ern{Lam12b} its follows that:
\beq
\Lambda_{12}^{12} = \Lambda_{21}^{21} = 0  \quad \Rightarrow \quad \vec \Lambda^{12}  =
 \vec \Lambda^{21}  = 0.
\label{Lam12bzero}
\enq
Thus according to \ern{momgfex2}:
\beq
\vec P (t) = \frac{\mu_0}{4 \pi} Q_1 Q_2 \dot \phi \vec \Gamma^{21} \sin \Delta \phi.
\label{momgfex3b}
\enq
Assuming $\Delta \phi = \frac{\pi}{2}$ and a long enough wire:
\beq
\vec P (t) =  - \frac{\mu_0}{4 \pi} Q_1 Q_2 \dot \phi \hat y.
\label{momgfex4}
\enq
Hence, such a system will gain momentum in the $y$ direction (perpendicular to the wire direction) depending on the phase temporal derivative. The total current flowing through the wire (say wire 1)  is:
\beq
I_1 (t) = \int \vec J_1 \cdot \hat x dy dz  = Q_1  \dot \phi \cos \phi_1,
\qquad -x_{01} < x  < +x_{01}
\label{curr1b}
\enq
If we take $\phi_1 = \omega t$:
\beq
I_1 (t) = Q_1  \omega \cos (\omega t)
\label{curr12}
\enq
which is the current that is needed to charge and discharge the ends of the wire. The amplitude
of the said current is:
\beq
I_{a 1} = Q_1  \omega.
\label{curr13}
\enq
From \ern{f12}, \ern{rhoexpsiex} and \ern{rh11rh22} it follows that the total charge in the wire is null because the charge accumulated in one end is equal and opposite to the other end with a maximal
value of $Q_1$. The same goes of course in the case of wire 2. Thus we may write the
momentum equation as:
\beq
\vec P (t) =  - \frac{\mu_0}{4 \pi \omega} I_{a 1} I_{a 2} \hat y.
\label{momgfex5}
\enq

\section{Discussion}

In this paper, we have shown that, in general, Newton's third law is not compatible with the principles of special relativity and the total force on a two charged body system is not zero. Still, momentum is conserved if one takes the field momentum into account, and the same is true for energy.

The main results of this paper are given by \ern{momgfex5}, which describe the total relativistic force for a specific configuration. Although more general configurations are allowed (see \ern{momgf}).

The result shows that the higher the current amplitude and counterintuitive the lower the frequency, the higher the momentum obtained. This is somewhat misleading because if the frequency is too low
one may risk an aerial discharge (see discussion in \cite{RY4}). However, neglecting this problem
we obtain for two wires carrying a 10 kiloampere current, with 1 Hz frequency a momentum of 1.6 kg meter/second.

We remark that an "antigravity" effect may be obtain by changing the phase quadratically in time,
this will cause a second derivative of phase that may cause a temporal first derivative of momentum, i.e. a force.

Obviously to conduct an experimental verification of the suggested relativistic engine is highly desirable in order to corroborate the ideas presented in the current paper this is also left as a task for the future.

\section{Conclusion}

To conclude we make a comparison between the relativistic motor and other types of electromagnetic engines. A photon engine will emit photons backwards and thus propel itself forward. It
may be a powerful laser or a radio-frequency cavity  \cite{White} (the only difference between
those too cases are the energy and momenta of a single photon).
To reach a momentum $p$ using a photon engine one needs an energy of $E_p=pc$ while for a relativistic engine an energy of $E \sim \frac{1}{2} p v  $ will suffice. The~ratio is $\frac{E_p}{E } = \frac{2 c}{ v}$ which is a huge number for non relativistic~speeds.

Of course standard electric cars to today (Tesla for example) can reach significant speed and momentum, but unlike the relativistic motor they need a road to push against, otherwise no motion is possible.

\end{document}